# Amino Acid Translocation through a Dual Nanopore Platform

Chih-Yuan Lin[1#], Pia Bhatia[1#], Alexandra Sofia Uy-Tioco[1,2], Kyril Kavetsky[1,2], Celia Morral[1],

Rachael Keneipp[1], Namrata Pradeep[1,3], Marija Drndić[1*]

*1Department of Physics and Astronomy, University of Pennsylvania, Philadelphia, Pennsylvania*

*19104, United States*

*2 Department of Materials Science and Engineering, University of Pennsylvania, Philadelphia,*

*Pennsylvania 19104, United States*

*3Department of Chemical and Biomolecular Engineering, University of Pennsylvania,*

*Philadelphia, Pennsylvania 19104, United States*

*# These authors contributed equally to the work.*

**Abstract**

We demonstrate a dual nanopore platform (DNP) containing a top 2D $MoS_2$ pore in series with a 3 to 5 nm thick SiN pore, vertically separated by 30 nm, with diameters of 1.0 and 3.0 nm, respectively. This platform enables independent probing of analytes by each pore, thereby providing complementary information. We measure translocations of single amino acids (AA) and evaluate current blockades recorded across the two pores upon applying voltage. Small diameters ensured tight passage of individual AAs through the nanopores and provided a good signal-to-noise ratio (RMS current noise of 16 $pA_{RMS}$ and SNR = 6). We focus on measurements of O-Phospho-L-tyrosine at 400 mV, demonstrating single amino acid detection and a good quantitative agreement with the calculated open pore and blocked currents. Based on these results, future device performance can benefit from slightly smaller pores, specifically the SiN pore, higher voltages and electrolyte concentration, and lower system noise.

There is a growing interest in protein analysis, fingerprinting and sequencing that would enable single-cell analysis. Nanopore-based efforts leverage the successes of DNA and RNA[1] in protein pore[2–9] sequencers that are now commercially available.[10–12] The emerging landscape of single-molecule protein sequencing, recently reviewed by Alfaro et al[13] and Lu et al.[14], includes single-molecule protein identification technologies alongside innovations in mass spectrometry and nanopore (NP) sensors. Biological pores have recently been used to analyze proteins towards protein fingerprinting and sequencing[14] and protein (MspA) pores were employed for the analysis of peptide-oligonucleotide chain conjugates,[15,16] demonstrating a sequence-dependent signal. Advances in solid-state nanopores have also been significant (Xue et al.[17] and others[1,4,18–22]). Benefits of solid-state materials may include the stability of substrates[23], speed of data acquisition[24–26], enzyme-free operation, integration of transverse readouts, and sophisticated on-chip architectures[27–29]. Solid-state NPs also allow higher measurement bandwidths.[24,25,30] To discriminate between a molecule's building blocks (such as individual amino acids), an NP sensor requires: (1) small-diameter and thin pores, and (2) sufficient temporal resolution. The high signal levels of solid-state nanopores[17] at high bandwidths may make sequencing possible, but several challenges remain such as reproducible nanopore fabrication.

Thin solid-state NPs in silicon nitride (SiN)[31–33] (down to 1-nm-thick[34]) and in 2D materials (e.g., graphene[35], 2D molybdenum disulfide, $MoS_2$)[35–37] may hold some advantages for peptide and protein analysis at higher bandwidths in the 1-10 MHz regime and harsher environments where robustness to chemicals, temperatures and pressures may be important. Experimental translocation studies with sub-10 nm thick SiN and 2D NPs are limited.[38–40] Several studies applied single SiN NPs to differentiate the diameter, shape, charge, diffusion and dipole moment of proteins both experimentally and via modeling.[41,42] Other strategies besides direct detection have been experimentally pursued, such as trapping and probing the protein inside the NP.[44] Molecular dynamics (MD) studies showed how proteins can translocate through solid state NP in an unfolded fashion with a translocation time increasing with the molecular weight[45] and dependent on the protein's folding state[46] and nanopore diameters.[40] Theoretical investigations of proteins with added polylysine tags through 2D $MoS_2$ NPs[47] motivate the use of tags to facilitate protein loading.

In this Letter, we demonstrate single amino acid detection in a dual solid-state nanopore platform (DNP), consisting of two ultrathin and small diameter pores in series. This platform uniquely enables the independent probing of unmodified analytes by each pore – thereby providing complementary information from *two* pores in a single shot. Because of this, the DNP platform may also be used to probe the electrophoretic transit time of an analyte as it travels from one pore to the next. Enabled by state-of-the-art transmission electron microscopy (TEM) and aberration corrected scanning TEM (AC-STEM) drilling techniques, here we fabricate 1.0 and 3.0 nm diameter pores with ~ 30 nm separation on a silicon chip with 5 µm of insulating oxide (**Figure 1**). This design resulted in a sufficiently high signal-to-noise ratio (SNR) for the detection of single O-Phospho-L-Tyrosine molecules. Compared to our previous work on coupled nanopores for single molecule detection,[48] the pore diameters are significantly reduced (see **Table S2** in the **Supporting Information**).

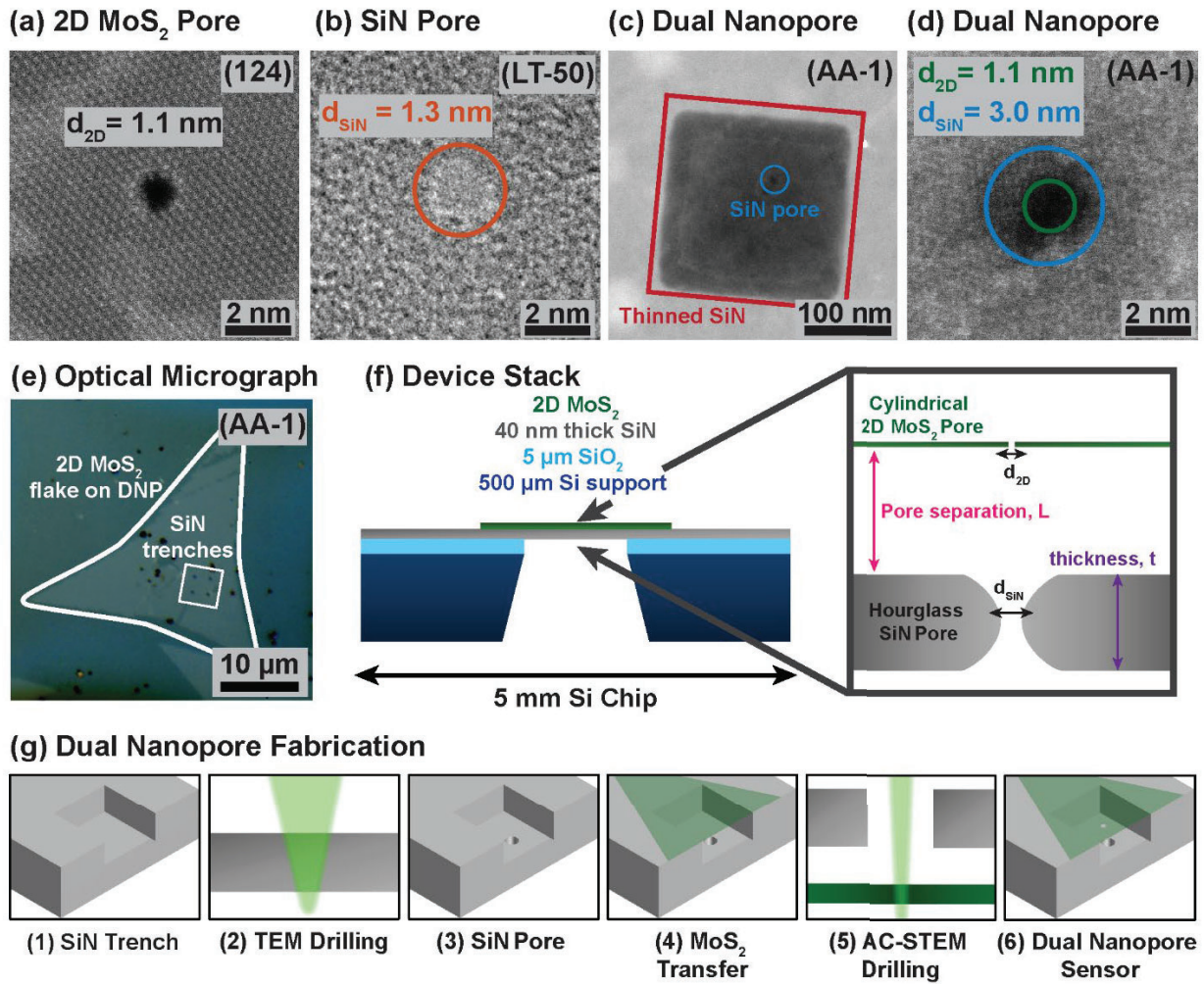

**Figure 1. Dual NP system for AA detection.** (a) AC-STEM image of a single 2D $MoS_2$ nanopore with 1.1 nm diameter, (b) TEM image of a single SiN pore with 1.3 nm diameter. (c, d) AC-STEM images of a DNP system showing the SiN trench and the drilled 2D pore. The red square shows the ~ 200 nm by 200 nm SiN trench on top of which the 2D flake is placed. The blue circle is the region where the two nanopores are drilled. The green circle indicates the 2D pore. (e) Optical micrograph showing the 2D $MoS_2$ flake placed on top of the SiN membrane containing four SiN trenches that can be used to make the DNPs. The 2D flake is placed on top of a trench etched into a supporting SiN membrane and both pores are drilled with a highly focused electron beam. (f) Cross-sectional view of the DNP chip with layers indicated. (g) Flowchart of DNP device fabrication steps.

**Figures 1a-c** demonstrate the capabilities of TEM and AC-STEM drilling for the fabrication of solid-state NPs and the fabrication of the DNP. In **Figures 1a** and **1b** we show a single pore in 2D MoS ($d_{2D}$ = 1.1 ± 0.01 nm) fabricated by AC-STEM drilling, and a single pore in SiN ($d_{SiN}$ =1.3 ± 0.04 nm) fabricated by TEM drilling. In **Figures 1c** and **1d,** we show TEM images of a fully integrated DNP device with a top 2D pore ($d_{2D}$ = 1.1 ± 0.11 nm) and a bottom SiN pore ($d_{SiN}$ = 3.0 ± 0.29 nm) fabricated using AC-STEM and TEM drilling, respectively. The SiN trench is within the red square in **Figure 1c**. The 2D pore is located within the green circle and the SiN pore is 30 nm under the 2D pore and indicated in **Figure 1d** within the blue circle. **Figure 1e** shows an optical image of a DNP device, with the freestanding SiN membrane and the MoS$_2$ flake outlined in white. The membrane contains four SiN trenches that can be used for DNP fabrication. The growth of 2D MoS$_2$ flake, the flake transfer details and the accurate measurements of pore diameters from TEM and AC-STEM images can be found in the **Supporting Information** (**Sections 2** and **3**).

**Figure 1f** illustrates a schematic of the cross-sectional view of the DNP chip with layers indicated. The 5-μm thick layer of SiO$_2$ serves to lower the capacitive noise. The right panel captures the geometries of the 2D MoS$_2$ and SiN pores, respectively; this detail is pertinent to subsequent discussions of conductance in the DNP. Finally, in **Figure 1g**, we summarize the fabrication process for our DNP platform. Beginning with a pre-patterned square trench ($\sim$ 200 × 200 nm$^2$), a JEOL F200 operating in TEM mode at 200 kV was used to drill the first pore in SiN (**Section 1** of **Supporting Information**). Then, a single 2D MoS$_2$ flake, grown via chemical vapor deposition, was transferred on top the SiN membrane, as detailed in **Section 2** of **Supporting Information**.[49] The 2D flakes are large enough to easily cover the trenches. Finally, a JEOL NEOARM AC-STEM operating at 80 kV was used to drill the top 2D pore in monolayer MoS$_2$.

The details of AC-STEM drilling down to the sub-nm range in 2D materials have been reported previously.[49–51]

Ion transport in DNP devices was modeled by modified Poisson-Nernst-Planck and Navier-Stokes (PNP-NS) equations with the associated boundary conditions. This continuum-based theory is valid for describing ion transport when the scale is larger than ~1 nm.[52] All simulations were carried out with *COMSOL Multiphysics (version 5.6)*. In line with the experiment, the simulated system contains a top 2D pore having a 1.1 nm diameter and a bottom pore having a 3 nm diameter, separated by 30 nm. A voltage bias of 400 mV is applied across the DNP. We assumed a surface charge of - 20 mC/m$^2$ for both SiN and MoS$_2$ pores, respectively, following values from the literature.[53,54] The amino acid is modeled as a charged rod with dimensions of 0.6 nm × 1.0 nm,[55] electrophoretically moving along the axial direction. Details of the modeling are provided in the **Supporting Information**. The simulated fractional blockade (**Figure 2a**) shows two distinct signals, corresponding to analyte translocations through the two pores. Since the two layers are sufficiently separated (L = 30 nm) and the amino acid is much smaller (approximately 1 nm) than the separation, the blockades produced by each pore are expected to appear independently, as reflected by the two peaks of the electric field (black curve of **Figure 2b**).

As shown in **Figure 2a**, the fractional current blockade expected from the thicker SiN pore (~ 0.8 %) is smaller than that from the ultrathin 2D pore (~ 8 %). These fractional blockades for the two pores in the DNP are smaller than the blockades that would be expected in a single 2D or a single SiN nanopore at the same applied voltage (~20 % for a 2D pore and ~3 % for a SiN pore, as shown in **Section 6** of **Supporting Information**). We attribute this to two factors: (i) In the DNP configuration, the electric field across each individual pore is lower than the electric field across a single pore (gray line) if the total voltage is fixed (**Figure 2b**). (ii) Due to coupling effects in the DNP, the ion concentration increases nonlinearly in the trench region when the analyte enters the 2D pore (see **Figure 2c**), partially compensating for ions that would supposedly be excluded by the analyte. Therefore, the level of blockade is reduced. This motivates our interest in how the dual nanopore configuration regulates analyte translocation signals. As the next step, we examine the blockades generated when the analyte is in each pore. We keep the 2D pore diameter fixed to 1.1 nm (same as device AA-1 in **Figure 1d**) while varying the diameter of the bottom pore, as the 2D pore acts as the primary sensing region and its diameter needs to be ~ 1 nm to achieve a sufficient SNR for detecting small molecules such as amino acids.

**Figure 2e** shows the fractional current blockade from individual pores in the DNP as a function of the SiN pore diameter. The total applied voltage is distributed across the two pores, and $|\Delta I/I|_{2D}$ and $|\Delta I/I|_{SiN}$ denote the fractional current blockades when the amino acid is located within the 2D and SiN pores, respectively. Specifically, if the diameter of the 2D pore is fixed ($d_{2D}$ = 1.1 nm) while increasing the SiN diameter, $|\Delta I/I|_{2D}$ increases whereas $|\Delta I/I|_{SiN}$ decreases greatly. This is because the resistance of the 2D pore increasingly exceeds that of the SiN pore, so the blockade becomes governed entirely by the 2D pore, and the system evolves into a single-pore scenario. **Figure S3** shows the ratio of the DNP vs. 2D pore conductance, $G_{DNP}/G_{2D}$, reaching $G_{DNP}/G_{2D}$ ~ 1 when $d_{SiN}$ ~ 10 nm. When the diameter of the SiN pore is smaller than 2 nm, the two pores have comparable resistances, therefore the magnitudes of $|\Delta I/I|_{2D}$ (~ 3.3 %) and $|\Delta I/I|_{SiN}$ (~ 2.7 %) reach similar values (shown in **Figure 2e** for $d_{SiN}$ = 2 nm). The corresponding blockade profile is presented in **Figure 2d**, revealing two signals having similar magnitudes. This is further supported by the electric field profile (blue curve of **Figure 2b**). In this scenario, it may be challenging to resolve from which of the two pores the signal is coming. The results indicate that engineering an appropriate geometry is essential in the DNP platform.

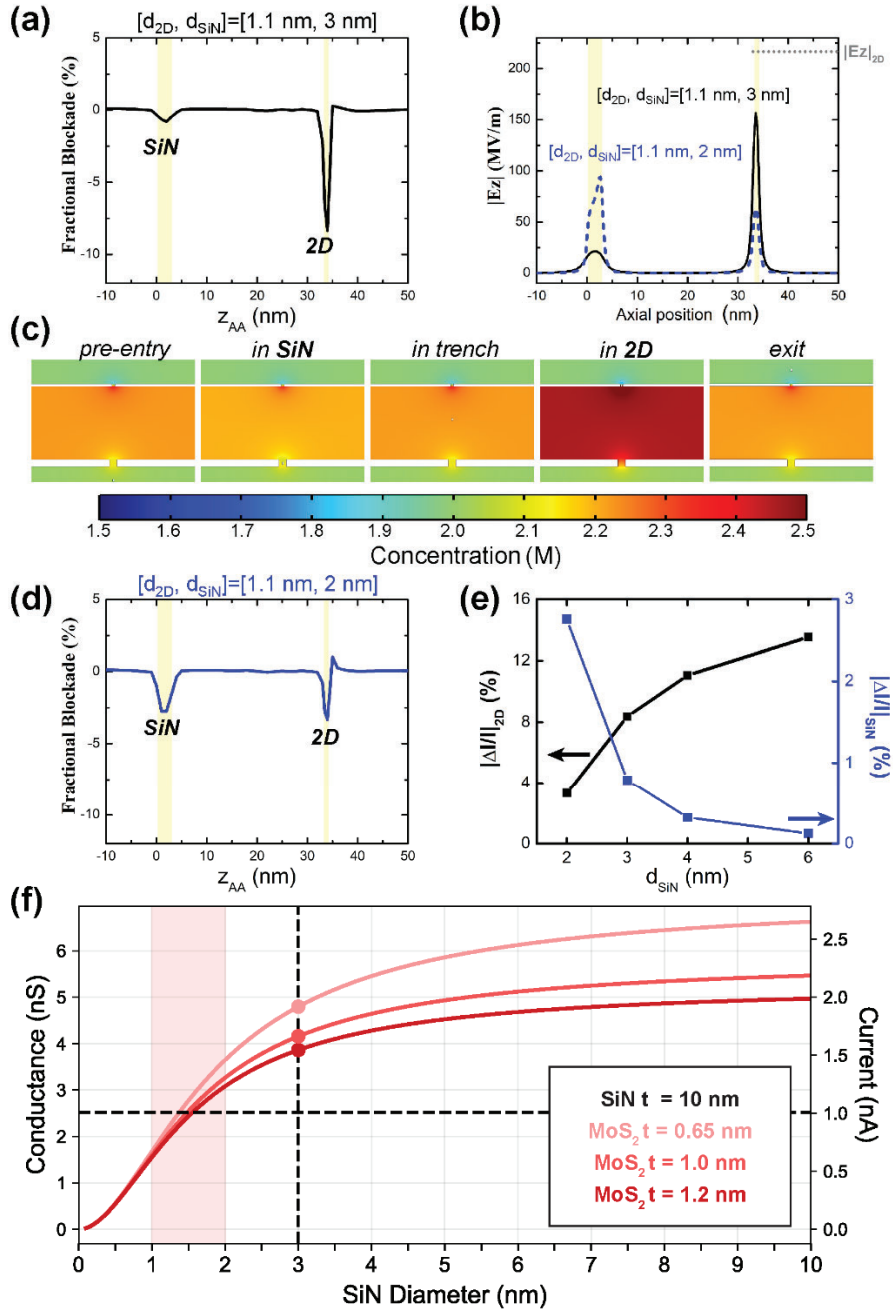

**Figure 2**. **Modeling of the dual nanopore platform (DNP).** (a,d) Simulated ion current blockade upon passage of amino acid through the DNP. $z_{AA}$ indicates the position of the amino acids of assumed size 0.6 nm by 1.0 nm through the two pores. (b) Axial variation of the strength of electric field in the DNP for different pore diameters. Black curve: $[d_{2D}, d_{SiN}] = [1.1 \text{ nm}, 3 \text{ nm}]$; blue curve: $[d_{2D}, d_{SiN}] = [1.1 \text{ nm}, 2 \text{ nm}]$. The thicknesses of 2D and SiN pores are 1 nm and 3 nm, respectively. The gray dashed line in the top right corner represents the electric field calculated from a single 2D pore with 1.1 nm in diameter and 1 nm in thickness at 400 mV across. (c) Concentration

(potassium and chloride ions) profiles representing different stages of amino acid translocation (i.e., $z_{AA}$). (e) Fractional blockades for individual pores in the DNP as a function of the SiN pore diameter, $d_{SiN}$. $|\Delta I/I|_{2D}$ and $|\Delta I/I|_{SiN}$ denote the blockades when an amino acid occupies the 2D and SiN pores, respectively. The voltage is 400 mV, and the electrolyte is 1 M KCl. Pore charges are also assumed as discussed in the text. (f) Conductance calculated based on two resistors in series (**eq 1**). SiN pore diameter varies with different MoS$_2$ thicknesses for 0.65 nm to 1.2 nm. The SiN thickness is fixed at 10 nm, that is, the effective thickness $t_{eff}$ =3.3 nm. The current is calculated for when the voltage is 400 mV.

To perform ion transport measurements, the 5 × 5 mm$^2$ DNP chips were sealed in a flow cell with silicone gaskets. Wetting of solid-state pores is challenging, particularly when the pores are very small.[49] To achieve wetting of the DNP, the chips are immersed in 1:1 ethanol/DI solution for at least 1 hour before ionic measurement. We then run I-V scans to obtain the value of ion conductance, which should be sufficiently high to ensure the pores are both wetted. If pores are not yet wetted, we continue sweeping voltages up to 500 mV while monitoring the conductance (**Section 7** of **Supporting Information** and **Figure 3b**). The inset of **Figure 3b** presents the wetting dynamic process monitored by I-V curves for the DNP device. Here, the gray curve shows a very small conductance (~ 0.11 nS). In contrast, wetted DNP displays a fairly linear I-V curve and a stabilized, significantly larger conductance (~ 2.52 nS) that is similar to but slightly smaller than the calculated value based on a simple resistor model, predicting the open DNP conductance, $G_{DNP}$, in a range from 3.9 nS to 4.8 nS assuming the effective thickness of the SiN pore 3.3 nm and the 2D pore from 0.65 to 1.2 nm, $d_{SiN}$ = 3.0 nm, and $d_{2D}$ = 1.1 nm (see the calculated values in **Figure 2f** where we plot the $G_{DNP}$ vs. $d_{SiN}$).

The dual nanopore sensor consists of two pores fabricated in series; therefore, from a simple resistor model, the conductance of the DNP can be estimated as[48]

$$G_{DNP} = \frac{G_{2D}G_{SiN}}{G_{2D}+G_{SiN}} \tag{eq 1}$$

, where $G_{DNP}$, $G_{2D}$ and $G_{SiN}$ are the conductances of the DNP, the 2D and the SiN pore, respectively. The discussion of the resistor model calculations is presented in **Section 5** in the **Supporting Information** and **Figure S2**. The results obtained from the numerical model show good agreement with the results based on the resistor model (**Figures 2f** and **S2**). This also means that the simple resistor model is applicable for these devices and can be used for rough estimates of the open pore current, useful for comparison with measured values. The increased noise in the low frequency regime compared to the unwet pore further supports that the pore is wetted (**Figure 3b**).

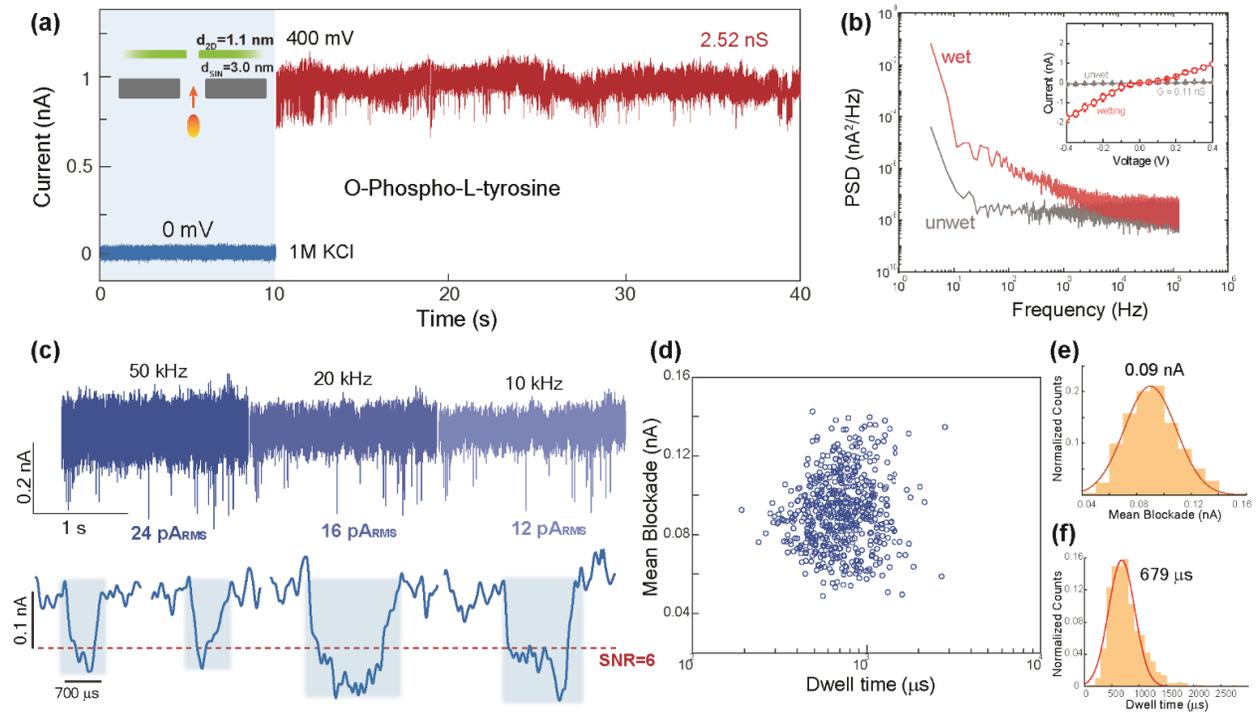

**Figure 3. O-Phospho-L-tyrosine translocation events in the DNP.** (a) Current trace at 400 mV and 1 M KCl, recorded at a sampling rate of 250 kHz. DNP geometries: $d_{2D} = 1.1$ nm; $d_{SiN} = 2.0$ nm; L (pore separation) = 30 nm. O-Phospho-L-tyrosine is translocated from the SiN to the 2D pore. (b) Noise power spectrum of wetted and unwetted devices. The inset shows the associated I-V curves. (c) A 2-second trace is filtered at the cutoff frequencies of 50 kHz, 20 kHz, and 10 kHz. Event examples of O-Phospho-L-tyrosine translocation. The red dashed line indicates SNR = 6. (d) Scatter plot of the mean event current blockades vs. event durations. Data corresponds to capture rate of 593 events per 30 second trace. Cuttoff frequency used is 20 kHz at SNR ~ 6. (e, f) Histograms of mean blockades, (e), and dwell times, (f), with Gaussian fitted curves.

**Figure 3** demonstrates the detection of a single amino acid through the DNP system with pore diameters $d_{SiN}$ = 3 nm, $d_{2D}$ = 1.1 nm, and pore separation of L ~ 30 nm (this device is shown in **Figure 1d**). Here, O-Phospho-L-tyrosine was chosen as a model amino acid analyte for our DNP study. Of the standard amino acids, tyrosine is considered large but is weakly charged.[56] Therefore, we hypothesized that the additional phosphate group on O-Phospho-L-tyrosine would enhance electrophoretic mobility of the molecule in our system. As illustrated in **Figure 3a**, the SiN side of the flow cell was filled with 1 M KCl containing the 2 μM O-Phospho-L-tyrosine and is grounded. The analyte solution was prepared with 1 M KCl and DI water without adding any buffers, to avoid getting signals from buffer ions, which is possible in our case because of the extremely small pore size. Thus, the solution pH was approximately 6. In the absence of an applied electric field (i.e., 0 mV), no translocation events were observed within the 10 second window (blue trace in **Figure 3a**). When a voltage of 400 mV is applied, the amino acid enters the SiN pore first and then the 2D pore, producing clear blockade signals (red trace in **Figure 3a**). Given the sub-nanometer dimensions of the analytes, the resulting current drop is expected to be small, as modeled in **Figure 2**. Therefore, achieving a sufficient SNR is essential to get visible translocation events.

**Figure 3b** showcases a 2-second current trace under different cutoff frequencies. Using low-noise chips with 5-μm thick $SiO_2$ layer underneath the SiN membrane yields noise levels as low as 24, 16, and 12 $pA_{RMS}$ at cutoff frequencies of 50 kHz, 20 kHz, and 10 kHz, respectively. Data analysis was performed by establishing a threshold search and obtaining histograms of currents and dwell times, as reported previously.[30,48] Events were then collected from the trace at a cutoff frequency of 20 kHz and the detection threshold was 0.1 nA, which corresponds to SNR = 6. The capture rate is approximately ~ 20 events/second at 400 mV.

We note that no events were observed over a 30 second time window when the voltage was < 200 mV. On the other hand, when the voltage was increased to 500 mV, the device was clogged shortly after biasing and could not be reopened again. As the trench can function as a pre-concentration region, lowering the analyte concentration may mitigate clogging issues in future measurements. The fact that we observe translocation events at 400 mV proves that the amino acid molecules, driven by the applied electric field, can travel through the DNP without getting stuck in the trench region under these conditions. Here the trench is about 10 times larger (200 by 200 nm) than the pore separation, 30 nm. In this case, the trench walls do not affect the field distribution (also see **Figure 2b**), and the DNP can be modeled as two pores in infinitely large parallel planes. Future work will explore the effects of trench sizes on the electric field profile, as an additional knob to tune DNP characteristics.

Representative translocation events are shown in **Figure 3c**. The scatter plot of mean current blockades vs. dwell times is shown in **Figure 3d**. The mean current blockades primarily falls in the range of 0.08 nA - 0.12 nA, corresponding to fractional blockades of approximately 8 % – 12 %. Intriguingly, this agrees well with our modeling, which predicts an ~ 8% blockade arising from the 2D pore in the DNP configuration with the same pore diameters and thicknesses (see **Figure 2a**). Our model predicted a much smaller mean fractional current blockade of ~ 0.8 % for the SiN pore. Thus, we attribute the detected events mainly to the signals from the 2D pore. Although filtering to a lower cutoff frequency may allow signals from the SiN pore to be resolved, this would not only distort event depth but also sacrifice time resolution, hindering detection of fast translocation events, which is undesirable. Furthermore, we observed that detected event durations fall within one order of magnitude, from $2 \times 10^2$ μs to $2 \times 10^3$ μs, revealing a narrow time distribution, in contrast to event durations for DNA translocation[30] which are typically distributed over a wider range. This is not surprising because the chosen analyte is a single molecule that is small (1 nm) and uniform in size. The mean current blockade and characteristic dwell time are ~ 0.09 nA and 679 μs, respectively (see **Figures 3e** and **3f**). The observed dwell times on the order of several hundred microseconds may be attributed to the drag force from electroosmotic flow arising from the negatively charged nanopore walls, which opposes the electrophoretic motion of the amino acid.[54,57,58]

We note that the fabrication of these DNP devices with small diameters remains challenging, But, as we have shown in this manuscript, fabrication is certainly possible with state-of-the-art tools. The yield of dual nanopore devices, defined as % of devices that survived all steps, from the fabrication to obtaining translocation data in this work is 13%. Failure modes included contamination during 2D nanopore drilling in the AC-STEM and failure of DNP devices to properly wet. Both of these challenges can be ameliorated in the future.

In conclusion, relying on exemplary advancements in aberration-corrected electron microscopy, we have made a solid-state system containing two small pores connected in series and vertically spaced by a small distance which can be tuned with a 1 nm precision either via RIE etching[31] or STEM thinning[34]. TEM and AC-STEM imaging also provided characterizations of the pores in terms of accurate measurements of diameters (with accuracy 0.1 nm or better), which informed and validated our experiment and modeling. Theoretically, this platform allows the measurement of at least five parameters: two sets of current blockades and event durations from the two pores, and the travel time of the analyte between the pores. Because the pores are also very thin (0.65 nm[37] and 3 nm, respectively in our experiments, **Figure 3**), the total resistance measured from the two-pore system, 0.4 GΩ, is small enough to allow for a high enough signal (0.1 nA) for single amino acid detection. Both nanopores were fabricated by state-of-the art TEM drilling capable of reaching the atomic scale. Given the imaging resolution of these advanced electron microscope techniques,[37] in principle, future devices could in principle feature even smaller characteristic dimensions at the atomic scale. Of the twenty standard amino acids, their lengths and widths range from ~ 0.4 to 1 nm and 0.3 to 0.6 nm, respectively.[55] Therefore, to detect individual amino acids with a solid-state NP, pore diameter and thickness must both be no greater than ~ 1 to 2 nm, and preferably < 1 nm for highest SNR. Fortunately, the resolution of AC-STEM has reached dimensions down to ~ 0.040 nm[59] making it the ideal technique for NP fabrication down to sub-nm scale.[49,51]

Future work can attempt to detect amino acids signals from both pores in the DNP and measure the transit time of a single amino acid between the two pores. This would provide a set of 5 parameters to characterize the amino acids (two current blockades, two dwell times and a transit time). From our experimental results (**Figure 3**) and modeling (**Figure 2**) we conclude that further device refinements to make smaller pores, particularly reducing the size of the SiN pore, possibly using higher voltages and higher electrolyte concentrations, and reducing the system's noise, appear necessary. The present resolution in aberration-corrected electron microscopy[59] allows ample room for further improvements. Single-layer sub-nm diameter 2D $MoS_2$ pores have been demonstrated[49] by AC-STEM drilling. While in this work we also used AC-STEM drilling to make the top 2D pore sufficiently small, the bottom SiN pore should be similarly made smaller in the future by reducing either the diameter[24] or the thickness (for example via STEM thinning)[34] or both. Our work demonstrates the first step towards highly-sensitive multiple-pore systems at the 1-nm-scale using 2D materials and locally-thinned[34,60] SiN membranes for direct analyte detection.

## Associated Content

The Supporting Information is available free of charge via the Internet at http://pubs.acs.org. Nanopore drilling (Section 1), 2D MoS2 growth and transfer (Section 2), characterization of pore size (Section 3), comparison with previous work (Section 4), resistor model (Section 5), numerical model (Section 6), ionic measurements and wetting procedure (Section 7).

## Author Information

*Corresponding Author(s)*

*Email: drndic@physics.upenn.edu

*Author Contributions*

C.Y.L, K.K., and M.D. conceived the experiment. C.Y.L performed electron beam lithography and reactive ion etching to fabricate locally thinned trenches. C.Y.L and K.K. performed TEM drilling to fabricate SiN pores. P.B. annealed devices and performed AC-STEM drilling to fabricate single $MoS_2$ pores and $MoS_2$ pores on DNP chips. A.S.U. and N. P. grew and transferred 2D $MoS_2$. A.S.U and R.N.K. performed focused ion beam milling of substrates for single $MoS_2$ pore devices. C.Y.L. performed ionic measurements, numerical modeling and data analysis. C.M. analyzed pore diameters from TEM and AC-STEM imaging. R.N.K. annealed devices and performed AC-STEM drilling to fabricate single $MoS_2$ pores. All authors discussed the manuscript and contributed to the writing.

## Acknowledgements

The work was supported by NIH grants R21HG012395 and R01HG012413. C.M. also acknowledges that this material is based upon work supported by the National Science Foundation Graduate Research Fellowship Program under Grant No(s) (DGE-2236662). R.N.K.

acknowledges the Vagelos Institute for Energy Science and Technology at the University of Pennsylvania for a graduate fellowship. Any opinions, findings, and conclusions or recommendations expressed in this material are those of the author(s) and do not necessarily reflect the views of the National Science Foundation. This work was carried out in part at the Singh Center for Nanotechnology, supported by the NSF grant NNCI-2025608. The authors acknowledge the use of facilities supported by the Laboratory for Research on the Structure of Matter and the NSF through the University of Pennsylvania Materials Research Science and Engineering Center (MRSEC) DMR-2309043. We also acknowledge Dr. Douglas Yates for useful discussions regarding AC-STEM imaging and drilling. We also acknowledge discussions with Prof. Ken Shepard and Prof. Michael Roukes about the potential use of dual nanopores for protein sequencing.

# References

(1) Branton, D.; Deamer, D. W.; Marziali, A.; Bayley, H.; Benner, S. A.; Butler, T.; Ventra, M. D.; Garaj, S.; Hibbs, A.; Huang, X.; Jovanovich, S. B.; Krstic, P. S.; Lindsay, S.; Ling, X. J.; Mastrangelo, C. H.; Meller, A.; Oliver, J. S.; Pershin, Y. V.; Ramsey, J. M.; Riehn, R.; Soni, G. V.; Tabard-Cossa, V.; Wanunu, M.; Wiggin, M.; Schloss, J. A. The Potential and Challenges of Nanopore Sequencing. *Nature Biotechnology* **2008**, *26* (10), 1146–1153. https://doi.org/10.1038/nbt.1495.

(2) Kasianowicz, J. J.; Brandin, E.; Branton, D.; Deamer, D. W. Characterization of Individual Polynucleotide Molecules Using a Membrane Channel. *PNAS* **1996**, *93* (24), 13770–13773. https://doi.org/10.1073/pnas.93.24.13770.

(3) Akeson, M.; Branton, D.; Kasianowicz, J. J.; Brandin, E.; Deamer, D. W. Microsecond Time-Scale Discrimination among Polycytidylic Acid, Polyadenylic Acid, and Polyuridylic Acid as Homopolymers or as Segments within Single RNA Molecules. *Biophysical Journal* **1999**, *77* (6), 3227–3233. https://doi.org/10.1016/S0006-3495(99)77153-5.

(4) Deamer, D. W.; Akeson, M. Nanopores and Nucleic Acids: Prospects for Ultrarapid Sequencing. *Trends in Biotechnology* **2000**, *18* (4), 147–151. https://doi.org/10.1016/S0167-7799(00)01426-8.

(5) Deamer, D. W.; Branton, D. Characterization of Nucleic Acids by Nanopore Analysis. *Acc. Chem. Res.* **2002**, *35* (10), 817–825. https://doi.org/10.1021/ar000138m.

(6) Cherf, G. M.; Lieberman, K. R.; Rashid, H.; Lam, C. E.; Karplus, K.; Akeson, M. Automated Forward and Reverse Ratcheting of DNA in a Nanopore at Five Angstrom Precision. *Nat. Biotechnol.* **2012**, *30* (4), 344–348. https://doi.org/10.1038/nbt.2147.

(7) Derrington, I. M.; Butler, T. Z.; Collins, M. D.; Manrao, E.; Pavlenok, M.; Niederweis, M.; Gundlach, J. H. Nanopore DNA Sequencing with MspA. *PNAS* **2010**, *107* (37), 16060–16065. https://doi.org/10.1073/pnas.1001831107.

(8) Laszlo, A. H.; Derrington, I. M.; Ross, B. C.; Brinkerhoff, H.; Adey, A.; Nova, I. C.; Craig, J. M.; Langford, K. W.; Samson, J. M.; Daza, R.; Doering, K.; Shendure, J.; Gundlach, J. H. Decoding Long Nanopore Sequencing Reads of Natural DNA. *Nature Biotechnology* **2014**, *32* (8), 829–833. https://doi.org/10.1038/nbt.2950.

(9) Manrao, E. A.; Derrington, I. M.; Laszlo, A. H.; Langford, K. W.; Hopper, M. K.; Gillgren, N.; Pavlenok, M.; Niederweis, M.; Gundlach, J. H. Reading DNA at Single-Nucleotide Resolution with a Mutant MspA Nanopore and Phi29 DNA Polymerase. *Nat Biotechnol* **2012**, *30* (4), 349–353. https://doi.org/10.1038/nbt.2171.

(10) Player, R.; Verratti, K.; Staab, A.; Bradburne, C.; Grady, S.; Goodwin, B.; Sozhamannan, S. Comparison of the Performance of an Amplicon Sequencing Assay Based on Oxford Nanopore Technology to Real-Time PCR Assays for Detecting Bacterial Biodefense Pathogens. *BMC Genomics* **2020**, *21* (1), 166. https://doi.org/10.1186/s12864-020-6557-5.

(11) Jain, M.; Koren, S.; Miga, K. H.; Quick, J.; Rand, A. C.; Sasani, T. A.; Tyson, J. R.; Beggs, A. D.; Dilthey, A. T.; Fiddes, I. T.; Malla, S.; Marriott, H.; Nieto, T.; O'Grady, J.; Olsen, H. E.; Pedersen, B. S.; Rhie, A.; Richardson, H.; Quinlan, A. R.; Snutch, T. P.; Tee, L.; Paten, B.; Phillippy, A. M.; Simpson, J. T.; Loman, N. J.; Loose, M. Nanopore Sequencing and Assembly of a Human Genome with Ultra-Long Reads. *Nature Biotechnology* **2018**, *36* (4), 338. https://doi.org/10.1038/nbt.4060.

(12) Wick, R. R.; Judd, L. M.; Holt, K. E. Deepbinner: Demultiplexing Barcoded Oxford Nanopore Reads with Deep Convolutional Neural Networks. *PLOS Computational Biology* **2018**, *14* (11), e1006583. https://doi.org/10.1371/journal.pcbi.1006583.

(13) Alfaro, J. A.; Bohländer, P.; Dai, M.; Filius, M.; Howard, C. J.; van Kooten, X. F.; Ohayon, S.; Pomorski, A.; Schmid, S.; Aksimentiev, A.; Anslyn, E. V.; Bedran, G.; Cao, C.; Chinappi, M.; Coyaud, E.; Dekker, C.; Dittmar, G.; Drachman, N.; Eelkema, R.; Goodlett, D.; Hentz, S.; Kalathiya, U.; Kelleher, N. L.; Kelly, R. T.; Kelman, Z.; Kim, S. H.; Kuster, B.; Rodriguez-Larrea, D.; Lindsay, S.; Maglia, G.; Marcotte, E. M.; Marino, J. P.; Masselon, C.; Mayer, M.; Samaras, P.; Sarthak, K.; Sepiashvili, L.; Stein, D.; Wanunu, M.; Wilhelm, M.; Yin, P.; Meller, A.; Joo, C. The Emerging Landscape of Single-Molecule Protein Sequencing Technologies. *Nat Methods* **2021**, *18* (6), 604–617. https://doi.org/10.1038/s41592-021-01143-1.

(14) Lu, C.; Bonini, A.; Viel, J. H.; Maglia, G. Towards Single-Molecule Protein Sequencing Using Nanopores. *Nature Biotechnology* **2025**, *43*, 312–322. https://doi.org/10.1038/s41587-025-02587-y.

(15) Yan, S.; Zhang, J.; Wang, Y.; Guo, W.; Zhang, S.; Liu, Y.; Cao, J.; Wang, Y.; Wang, L.; Ma, F.; Zhang, P.; Chen, H.-Y.; Huang, S. Single Molecule Ratcheting Motion of Peptides in a Mycobacterium Smegmatis Porin A (MspA) Nanopore. *Nano Lett.* **2021**, *21* (15), 6703–6710. https://doi.org/10.1021/acs.nanolett.1c02371.

(16) Brinkerhoff, H.; Kang, A. S. W.; Liu, J.; Aksimentiev, A.; Dekker, C. *Infinite Re-Reading of Single Proteins at Single-Amino-Acid Resolution Using Nanopore Sequencing*; preprint; Biophysics, 2021. https://doi.org/10.1101/2021.07.13.452225.

(17) Xue, L.; Yamazaki, H.; Ren, R.; Wanunu, M.; Ivanov, A. P.; Edel, J. B. Solid-State Nanopore Sensors. *Nature Reviews Materials* **2020**, *5* (12), 931–951. https://doi.org/10.1038/s41578-020-0229-6.

(18) Healy, K.; Schiedt, B.; Morrison, A. P. Solid-State Nanopore Technologies for Nanopore-Based DNA Analysis. *Nanomedicine (London, England)* **2007**, *2* (6), 875–897. https://doi.org/10.2217/17435889.2.6.875.

(19) Wanunu, M. Nanopores: A Journey towards DNA Sequencing. *Physics of Life Reviews* **2012**, *9* (2), 125–158. https://doi.org/10.1016/j.plrev.2012.05.010.

(20) Deamer, D.; Akeson, M.; Branton, D. Three Decades of Nanopore Sequencing. *Nat. Biotechnol.* **2016**, *34* (5), 518–524. https://doi.org/10.1038/nbt.3423.

(21) Danda, G.; Drndić, M. Two-Dimensional Nanopores and Nanoporous Membranes for Ion and Molecule Transport. *Current Opinion in Biotechnology* **2019**, *55*, 124–133. https://doi.org/10.1016/j.copbio.2018.09.002.

(22) Venkatesan, B. M.; Bashir, R. Nanopore Sensors for Nucleic Acid Analysis. *Nature Nanotechnology* **2011**, *6* (10), 615–624. https://doi.org/10.1038/nnano.2011.129.

(23) Chou, Y.-C.; Masih Das, P.; Monos, D. S.; Drndić, M. Lifetime and Stability of Silicon Nitride Nanopores and Nanopore Arrays for Ionic Measurements. *ACS Nano* **2020**, *14* (6), 6715–6728. https://doi.org/10.1021/acsnano.9b09964.

(24) Chien, C.-C.; Shekar, S.; Niedzwiecki, D. J.; Shepard, K. L.; Drndić, M. Single-Stranded DNA Translocation Recordings through Solid-State Nanopores on Glass Chips at 10 MHz Measurement Bandwidth. *ACS Nano* **2019**, *13* (9), 10545–10554. https://doi.org/10.1021/acsnano.9b04626.

(25) Rosenstein, J. K.; Wanunu, M.; Merchant, C. A.; Drndic, M.; Shepard, K. L. Integrated Nanopore Sensing Platform with Sub-Microsecond Temporal Resolution. *Nat. Methods* **2012**, *9* (5), 487–492. https://doi.org/10.1038/nmeth.1932.

(26) Shekar, S.; Niedzwiecki, D. J.; Chien, C.-C.; Ong, P.; Fleischer, D. A.; Lin, J.; Rosenstein, J. K.; Drndić, M.; Shepard, K. L. Measurement of DNA Translocation Dynamics in a Solid-State Nanopore at 100 Ns Temporal Resolution. *Nano Letters* **2016**, *16* (7), 4483–4489. https://doi.org/10.1021/acs.nanolett.6b01661.

(27) Puster, M.; Rodríguez-Manzo, J. A.; Balan, A.; Drndić, M. Toward Sensitive Graphene Nanoribbon–Nanopore Devices by Preventing Electron Beam-Induced Damage. *ACS Nano* **2013**, *7* (12), 11283–11289. https://doi.org/10.1021/nn405112m.

(28) Puster, M.; Balan, A.; Rodríguez-Manzo, J. A.; Danda, G.; Ahn, J.-H.; Parkin, W.; Drndić, M. Cross-Talk Between Ionic and Nanoribbon Current Signals in Graphene Nanoribbon-Nanopore Sensors for Single-Molecule Detection. *Small* **2015**, *11* (47), 6309–6316. https://doi.org/10.1002/smll.201502134.

(29) Healy, K.; Ray, V.; Willis, L. J.; Peterman, N.; Bartel, J.; Drndić, M. Fabrication and Characterization of Nanopores with Insulated Transverse Nanoelectrodes for DNA Sensing in Salt Solution. *Electrophoresis* **2012**, *33* (23), 3488–3496. https://doi.org/10.1002/elps.201200350.

(30) Lin, C.-Y.; Fotis, R.; Xia, Z.; Kavetsky, K.; Chou, Y.-C.; Niedzwiecki, D. J.; Biondi, M.; Thei, F.; Drndić, M. Ultrafast Polymer Dynamics through a Nanopore. *Nano Lett.* **2022**, *22* (21), 8719–8727. https://doi.org/10.1021/acs.nanolett.2c03546.

(31) Venta, K.; Shemer, G.; Puster, M.; Rodríguez-Manzo, J. A.; Balan, A.; Rosenstein, J. K.; Shepard, K.; Drndić, M. Differentiation of Short, Single-Stranded DNA Homopolymers in Solid-State Nanopores. *ACS Nano* **2013**, *7* (5), 4629–4636. https://doi.org/10.1021/nn4014388.

(32) Wanunu, M.; Dadosh, T.; Ray, V.; Jin, J.; McReynolds, L.; Drndić, M. Rapid Electronic Detection of Probe-Specific microRNAs Using Thin Nanopore Sensors. *Nature Nanotechnology* **2010**, *5* (11), 807–814. https://doi.org/10.1038/nnano.2010.202.

(33) Talaga, D. S.; Li, J. Single-Molecule Protein Unfolding in Solid State Nanopores. *J. Am. Chem. Soc.* **2009**, *131* (26), 9287–9297. https://doi.org/10.1021/ja901088b.

(34) Rodríguez-Manzo, J. A.; Puster, M.; Nicolaï, A.; Meunier, V.; Drndić, M. DNA Translocation in Nanometer Thick Silicon Nanopores. *ACS Nano* **2015**, *9* (6), 6555–6564. https://doi.org/10.1021/acsnano.5b02531.

(35) Merchant, C. A.; Healy, K.; Wanunu, M.; Ray, V.; Peterman, N.; Bartel, J.; Fischbein, M. D.; Venta, K.; Luo, Z.; Johnson, A. T. C.; Drndić, M. DNA Translocation through Graphene Nanopores. *Nano Lett.* **2010**, *10* (8), 2915–2921. https://doi.org/10.1021/nl101046t.

(36) Barrios-Perez, M. D.; Nicolai, A.; Delarue, P.; Meunier, V.; Drndic, M.; Senet, P. Improved Model of Ionic Transport in 2D MoS2 Membranes with Sub-5 Nm Pores. *Applied Physics Letters* **2019**, *114*, 023107. https://doi.org/DOI:%252010.1063/1.5061825.

(37) Liu, K.; Feng, J.; Kis, A.; Radenovic, A. Atomically Thin Molybdenum Disulfide Nanopores with High Sensitivity for DNA Translocation. *ACS Nano* **2014**, *8* (3), 2504–2511. https://doi.org/10.1021/nn406102h.

(38) Niedzwiecki, D. J.; Lanci, C. J.; Shemer, G.; Cheng, P. S.; Saven, J. G.; Drndić, M. Observing Changes in the Structure and Oligomerization State of a Helical Protein Dimer Using Solid-State Nanopores. *ACS Nano* **2015**, *9* (9), 8907–8915. https://doi.org/10.1021/acsnano.5b02714.

(39) Larkin, J.; Henley, R. Y.; Muthukumar, M.; Rosenstein, J. K.; Wanunu, M. High-Bandwidth Protein Analysis Using Solid-State Nanopores. *Biophysical Journal* **2014**, *106* (3), 696–704. https://doi.org/10.1016/j.bpj.2013.12.025.

(40) Tripathi, P.; Benabbas, A.; Mehrafrooz, B.; Yamazaki, H.; Aksimentiev, A.; Champion, P. M.; Wanunu, M. Electrical Unfolding of Cytochrome *c* during Translocation through a Nanopore Constriction. *Proc. Natl. Acad. Sci. U.S.A.* **2021**, *118* (17), e2016262118. https://doi.org/10.1073/pnas.2016262118.

(41) Yusko, E. C.; Bruhn, B. R.; Eggenberger, O. M.; Houghtaling, J.; Rollings, R. C.; Walsh, N. C.; Nandivada, S.; Pindrus, M.; Hall, A. R.; Sept, D.; Li, J.; Kalonia, D. S.; Mayer, M. Real-Time Shape Approximation and Fingerprinting of Single Proteins Using a Nanopore. *Nature Nanotechnology* **2017**, *12* (4), 360–367. https://doi.org/10.1038/nnano.2016.267.

(42) Houghtaling, J.; Ying, C.; Eggenberger, O. M.; Fennouri, A.; Nandivada, S.; Acharjee, M.; Li, J.; Hall, A. R.; Mayer, M. Estimation of Shape, Volume, and Dipole Moment of Individual Proteins Freely Transiting a Synthetic Nanopore. *ACS Nano* **2019**, *13* (5), 5231–5242. https://doi.org/10.1021/acsnano.8b09555.

(43) Ratinho, L.; Meyer, N.; Greive, S.; Cressiot, B.; Pelta, J. Nanopore Sensing of Protein and Peptide Conformation for Point-of-Care Applications. *Nat Commun* **2025**, *16* (1), 3211. https://doi.org/10.1038/s41467-025-58509-8.

(44) Schmid, S.; Stömmer, P.; Dietz, H.; Dekker, C. Nanopore Electro-Osmotic Trap for the Label-Free Study of Single Proteins and Their Conformations. *Nat. Nanotechnol.* **2021**, *16* (11), 1244–1250. https://doi.org/10.1038/s41565-021-00958-5.

(45) Restrepo-Pérez, L.; John, S.; Aksimentiev, A.; Joo, C.; Dekker, C. SDS-Assisted Protein Transport through Solid-State Nanopores. *Nanoscale* **2017**, *9* (32), 11685–11693. https://doi.org/10.1039/C7NR02450A.

(46) Li, J.; Fologea, D.; Rollings, R.; Ledden, B. Characterization of Protein Unfolding with Solid-State Nanopores. *Protein Pept Lett* **2014**, *21* (3), 256–265. https://doi.org/10.2174/09298665113209990077.

(47) Nicolaï, A.; Barrios Pérez, M. D.; Delarue, P.; Meunier, V.; Drndić, M.; Senet, P. Molecular Dynamics Investigation of Polylysine Peptide Translocation through MoS 2 Nanopores. *J. Phys. Chem. B* **2019**, *123* (10), 2342–2353. https://doi.org/10.1021/acs.jpcb.8b10634.

(48) Chou, Y.-C.; Lin, C.-Y.; Castan, A.; Chen, J.; Keneipp, R.; Yasini, P.; Monos, D.; Drndić, M. Coupled Nanopores for Single-Molecule Detection. *Nat. Nanotechnol.* **2024**, *19* (11), 1686–1692. https://doi.org/10.1038/s41565-024-01746-7.

(49) Thiruraman, J. P.; Masih Das, P.; Drndic, M. Stochastic Ionic Transport in Single Atomic Zero-D Pores. *ACS Nano* **2020**, *14* (19), 11831–11845. https://doi.org/10.1021/acsnano.0c04716.

(50) Masih Das, P.; Thiruraman, J. P.; Zhao, M.; Mandyam, S.; Johnson, A. T. C.; Drndic, M. Atomic-Scale Patterning in Two-Dimensional van Der Waals Superlattices. *Nanotechnology* **2019**, *31* (10), 105302 (7pp). https://doi.org/10.1088/1361-6528/ab596c.

(51) Keneipp, R. N.; Gusdorff, J. A.; Bhatia, P.; Shin, T. T.; Bassett, L. C.; Drndić, M. Nanoscale Sculpting of Hexagonal Boron Nitride with an Electron Beam. *The Journal of Physical Chemistry C* **2024**, *128* (21), 8741–8749. https://doi.org/10.1021/acs.jpcc.4c02038.

(52) Aluru, N. R.; Aydin, F.; Bazant, M. Z.; Blankschtein, D.; Brozena, A. H.; De Souza, J. P.; Elimelech, M.; Faucher, S.; Fourkas, J. T.; Koman, V. B.; Kuehne, M.; Kulik, H. J.; Li, H.-K.; Li, Y.; Li, Z.; Majumdar, A.; Martis, J.; Misra, R. P.; Noy, A.; Pham, T. A.; Qu, H.; Rayabharam, A.; Reed, M. A.; Ritt, C. L.; Schwegler, E.; Siwy, Z.; Strano, M. S.; Wang, Y.;

Yao, Y.-C.; Zhan, C.; Zhang, Z. Fluids and Electrolytes under Confinement in Single-Digit Nanopores. *Chem. Rev.* **2023**, *123* (6), 2737–2831. https://doi.org/10.1021/acs.chemrev.2c00155.

(53) Bandara, Y. M. N. D. Y.; Karawdeniya, B. I.; Hagan, J. T.; Chevalier, R. B.; Dwyer, J. R. Chemically Functionalizing Controlled Dielectric Breakdown Silicon Nitride Nanopores by Direct Photohydrosilylation. *ACS Appl. Mater. Interfaces* **2019**, *11* (33), 30411–30420. https://doi.org/10.1021/acsami.9b08004.

(54) Lin, C.-Y.; Turker Acar, E.; Polster, J. W.; Lin, K.; Hsu, J.-P.; Siwy, Z. S. Modulation of Charge Density and Charge Polarity of Nanopore Wall by Salt Gradient and Voltage. *ACS Nano* **2019**, *13* (9), 9868–9879. https://doi.org/10.1021/acsnano.9b01357.

(55) Ching, C. B.; Hidajat, K.; Uddin, M. S. Evaluation of Equilibrium and Kinetic Parameters of Smaller Molecular Size Amino Acids on KX Zeolite Crystals via Liquid Chromatographic Techniques. *Separation Science and Technology* **1989**, *24* (7–8), 581–597. https://doi.org/10.1080/01496398908049793.

(56) Pommié, C.; Levadoux, S.; Sabatier, R.; Lefranc, G.; Lefranc, M. IMGT Standardized Criteria for Statistical Analysis of Immunoglobulin V-REGION Amino Acid Properties. *J of Molecular Recognition* **2004**, *17* (1), 17–32. https://doi.org/10.1002/jmr.647.

(57) Sauciuc, A.; Morozzo Della Rocca, B.; Tadema, M. J.; Chinappi, M.; Maglia, G. Translocation of Linearized Full-Length Proteins through an Engineered Nanopore under Opposing Electrophoretic Force. *Nat Biotechnol* **2024**, *42* (8), 1275–1281. https://doi.org/10.1038/s41587-023-01954-x.

(58) Firnkes, M.; Pedone, D.; Knezevic, J.; Döblinger, M.; Rant, U. Electrically Facilitated Translocations of Proteins through Silicon Nitride Nanopores: Conjoint and Competitive Action of Diffusion, Electrophoresis, and Electroosmosis. *Nano Lett.* **2010**, *10* (6), 2162–2167. https://doi.org/10.1021/nl100861c.

(59) Chen, Z.; Jiang, Y.; Shao, Y.-T.; Holtz, M. E.; Odstrčil, M.; Guizar-Sicairos, M.; Hanke, I.; Ganschow, S.; Schlom, D. G.; Muller, D. A. Electron Ptychography Achieves Atomic-Resolution Limits Set by Lattice Vibrations. *Science* **2021**, *372* (6544), 826–831. https://doi.org/10.1126/science.abg2533.

(60) Drndic, M.; Wanunu, M.; Dadosh, T. High-Resolution Analysis Devices and Related Methods. US Patent 9121823, September 1, 2015. https://patents.justia.com/patent/9121823.

For Table of Contents Only

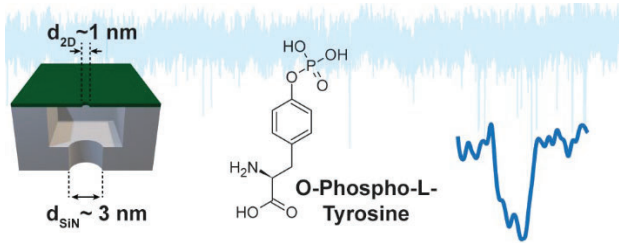

# Supplementary Information

## Amino Acid Translocation through a Dual Nanopore Platform

Chih-Yuan Lin[1], Pia Bhatia[1], Alexandra Sofia Uy-Tioco[1,2], Kyril Kavetsky[1,2], Celia Morral[1],

Rachael Keneipp[1], Namrata Pradeep[1,3], Marija Drndić[1]

[1]*Department of Physics and Astronomy, University of Pennsylvania, Philadelphia, Pennsylvania 19104,*

*United States*

[2]*Department of Materials Science and Engineering, University of Pennsylvania, Philadelphia,*

*Pennsylvania 19104, United States*

[3]*Department of Chemical and Biomolecular Engineering, University of Pennsylvania, Philadelphia,*

*Pennsylvania 19104, United States*

**Table of Contents**

## 1. TEM and AC-STEM Drilling

In this study, JEOL F200 TEM and JEOL NEOARM AC-STEM were employed to drill pores in thin SiN and 2D $MoS_2$, respectively. We refer to these two approaches as "TEM drilling" and "AC-STEM" drilling, respectively. The JEOL F200 was operated at 200 kV and drilling was performed by condensing the electron beam (~ 20 nA) to form a Ronchigram on the SiN region of interest at ~ 1 million times magnification. We monitored the Ronchigram on the fluorescent screen until the edge of the pore was just visible. Once a pore was opened, the electron beam was expanded immediately and an image was captured.

$MoS_2$ samples were thermally annealed in an $Ar/H_2$ environment for 1 to 2 hours prior to drilling. A 15-minute electron beam shower was also performed inside the microscope column. Drilling was carried out at 80 kV by parking the electron beam (~ 40 pA) at a particular location on the suspended $MoS_2$ at ~ 15 million times magnification. Once opened, we quickly captured an image of the $MoS_2$ pore.

## 2. 2D MoS$_2$ Growth & Device Integration

Single-layer MoS$_2$ flakes were grown via chemical vapor deposition (CVD). 9 mM ammonium heptamolybdate tetrahydrate and 23 mM sodium cholate solutions were prepared with deionized water and sonicated for one hour at room temperature. A piranha-cleaned Si/SiO$_2$ substrate (300 nm SiO$_2$) was spin-coated with both solutions and placed at the center of a 1-inch tubular furnace (Thermo Scientific Lindberg/Blue M). A 150 mg sulfur pellet, adhered to another piranha-cleaned Si/SiO$_2$ substrate, was placed 22 cm upstream. Prior to growth, the furnace was flushed with ultrapure N$_2$ gas at 1000 sccm for 10 minutes. Then, under 400 sccm flow, the furnace temperature was ramped up at a rate of 70 °C/min and held for 15 minutes at 750 °C, while the sulfur was simultaneously heated to and held at 180 °C. Afterwards, the furnace was turned off and left open for the samples to cool.

MoS$_2$ flakes were then transferred onto piranha-cleaned SiN windows via wet transfer. The flake-covered Si/SiO$_2$ substrate was spin-coated with A2 PMMA and placed face-up in 1.4 M KOH solution to etch the substrate off, leaving the flakes adhered to floating PMMA films. The films were transferred to a deionized water bath before being transferred onto a device and aligned under the optical microscope such that an MoS$_2$ flake completely covered the patterned holes. The device was then dried in air for an hour, then placed in a 60 °C acetone bath for 24 hours to remove the PMMA. Upon removal from acetone, the device was gently submerged in IPA for 30 seconds and was later subject to rapid thermal annealing (RTA) in a mixture of Ar and H$_2$ gases.

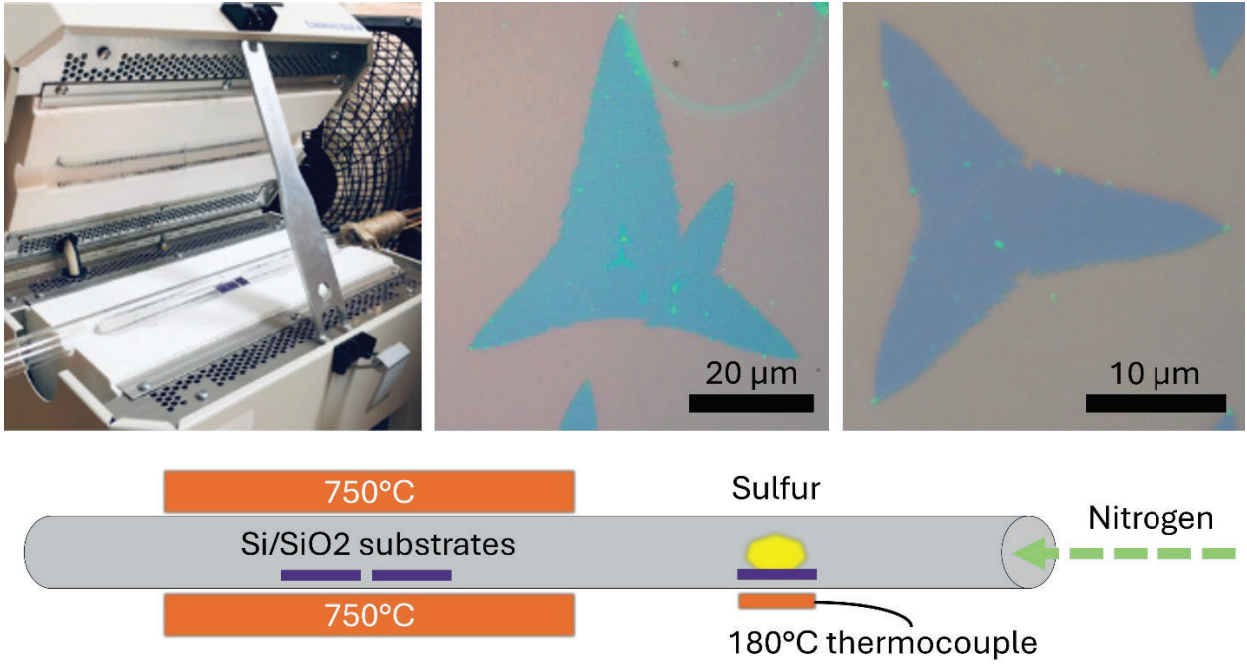

**Figure S1.** Schematic of chemical vapor deposition (CVD) growth of MoS$_2$, and optical micrographs of several 2D flakes.

**y**

### 3. Determining Pore Diameters from TEM and AC-STEM Images

Effective pore diameters were obtained from TEM images and AC-STEM images of SiN and $MoS_2$ pores, respectively. In both cases, the effective diameter was found by capturing the area of the pore and calculating the corresponding effective diameter, assuming a perfectly circular pore. To obtain the diameters of pores from TEM images, we averaged the pore area 5 times in ImageJ. The standard deviation is reported as the error. To obtain the sizes of pores from AC-STEM images, the area of the pore was calculated using Scipy ndimage and OpenCV functions to find the largest connected component of the image (*i.e.*, the pore) below a user-input threshold of pixel intensity. The optimal threshold was determined for each pore by visually comparing the highlighted pore area outputted by the code. To make the measurement more objective and reduce variability, we also calculated the pore area using a threshold 1% higher and lower than our chosen optimal threshold. *The effective pore diameter we report is the average of these three pore diameter outputs and the standard deviation is taken to be the error.* **Table S1** summarizes the effective diameters of the devices included in this study.

**Table S1. Effective Pore Diameter**

| Device | Effective SiN Diameter (nm) | Effective MoS₂ Diameter (nm) |
|---|---|---|
| AA-1 | $3.0 \pm 0.29$ | $1.1 \pm 0.11$ |
| MoS₂ Chip 124* | _ | $1.1 \pm 0.01$ |
| LT-50* | $1.3 \pm 0.04$ | _ |

*Single pore devices shown in **Figures 1a** and **1b** in the main text.

## 4. Comparisons with Previous Work on Coupled Nanopores

Previously, our group developed a similar two pore system (but larger pore diameters than in this work) to detect translocation of short DNA molecules.[1] We compare the devices with similar configuration that contains one 2D pore and one SiN pore, as listed in **Table S2**.

**Table S2. Comparisons of Pore Size with Previous Work**

| Device | Effective SiN Diameter (nm) | SiN thickness (nm) | Effective SiN thickness (nm)** | Effective MoS₂ Diameter (nm) | Pore separation, L (nm) |
|---|---|---|---|---|---|
| *This work* | | | | | |
| AA-1 | **3.0** | ~ 10 – 15 | ~ 3.3 – 5.0 | **1.1** | 30 |
| *Previous work (Chou et al.)*[1] | | | | | |
| L | 14.0 | 20 | 6.6 | 2.8 | 20 |
| M | 8.4 | 10 | 3.3 | 2.0 | 30 |
| N | 11.7 | 30 | 10 | 2.0 | 20 |

**\*\*** TEM drilling results in hourglass-shaped SiN pores.[2] In this case, the effective thickness is $\frac{1}{3}$ the total membrane thickness.

## 5. Dual Nanopore as Two Resistors in Series

The resistor-based model is used to estimate the ionic conductance through a nanopore, $G$, as follows

$$G = \frac{I}{V} = \sigma \left( \frac{4\,t_{eff}}{\pi d^2} + \frac{1}{d} \right)^{-1} \tag{1}$$

Here, $\sigma$ is the electrolyte conductivity, d is the pore diameter, $t_{eff}$ is the effective membrane thickness. Open pore current, I, is then given by G times the applied voltage, V. For 1 M KCl at room temperature, $\sigma \sim 11.5$ S/m; for cylindrical pores, $t_{eff}$ is equivalent to the pore thickness; and for hourglass-shaped pores (which result from TEM drilling), $t_{eff}$ is equivalent to one-third of the membrane thickness.[3,4] We treat the 2D $MoS_2$ pores as cylindrical, with fixed thickness t = 0.65 nm,[5] and the SiN pores as hourglass-shaped with variable thickness. The dual nanopore sensor consists of two pores fabricated in series (because the pore separation is much larger than the pore diameters); therefore, the conductance of the DNP can be estimated as[1]

$$G_{DNP} = \frac{G_{2D} G_{SiN}}{G_{2D} + G_{SiN}} \tag{2}$$

**Table S3** summarizes the conductance and open pore current for various dual nanopore geometries.

### Table S3. Dual Nanopore Conductance using resistor model

| Geometry | Individual Conductance | Dual Nanopore Conductance, $G_{DNP}$ | Open Pore Current at 400 mV |
|---|---|---|---|
| $MoS_2$ dia. = 1.1 nm<br>SiN dia. = 3.0 nm<br>SiN thickness = 10 nm | $G_{MoS2}$ = 7.2 nS<br>$G_{SiN}$ = 14.3 nS | 4.8 nS | 1.9 nA |

| MoS₂ dia. = 1.1 nm<br>SiN dia. = 3.0 nm<br>SiN thickness = 15 nm | $G_{MoS2}$ = 7.2 nS<br>$G_{SiN}$ = 11.1 nS | 4.4 nS | 1. 7 nA |
|---|---|---|---|

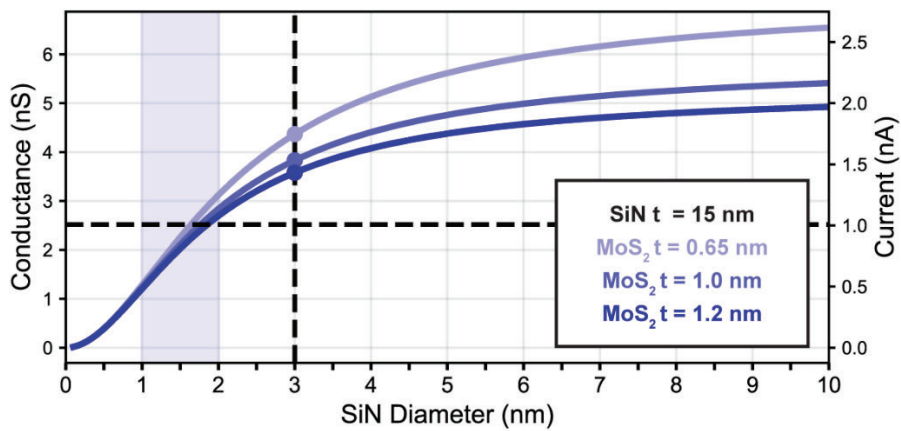

**Figure S2.** Conductance of the dual nanopore platform (DNP) as a function of SiN pore diameter for different $MoS_2$ thicknesses for 0.65 nm to 1.2 nm. The SiN membrane thickness is fixed at 15 nm, that is, the effective SiN pore thickness $t_{eff}$ = 5 nm. The ionic current is calculated for applied voltage of 400 mV.

## 6. Numerical Modeling

The simulated system is consistent with the device's geometry used in experiments (**Figure 1**): 2D layer and SiN layer are separated by L = 30 nm. A voltage bias is applied from the reservoir of 2D layer side while that of the SiN side is grounded. The diameter and thickness of 2D pore is fixed at 1.1 nm and 1 nm, respectively. The diameter of SiN is varied as discussed in the main text. The surfaces of both the 2D and SiN pores are negatively charged at neutral pH, and for simplicity, are assumed to carry the same charge density of −0.02 C/m$^2$.[6,7] To capture the underlying physics of the sub-nanometer-scale dual nanopore system, we employ the modified Poisson–Nernst–Planck (PNP) for electrokinetic ion transport,

$$-\varepsilon_f \nabla^2 \phi = F \sum_{i=1}^{2} z_i c_i \tag{3}$$

$$\nabla \cdot \left[ -D_i \nabla c_i - D_i \frac{z_i F}{RT} c_i \nabla \phi - a_i^3 D_i \frac{c_i \sum_{i=1}^{2} \nabla c_i}{1 - \sum_{i=1}^{2} a_i^3 c_i} + \mathbf{u} c_i \right] = 0 \tag{4}$$

In the above equation, $\phi$ is electric potential; $\mathbf{u}$ is the fluid velocity; $c_i$, $D_i$, and $z_i$ are the concentration, diffusivity, and valence of $i^{th}$ ionic species, respectively. Other physical constants include fluid permittivity, $\varepsilon_f$, Faraday constant, $F$, gas constant, $R$, and absolute temperature, $T$. Since the steric effect becomes significant at sub-nanometer scale, a correction term is introduced in the ionic flux, where $a_i$ denotes the effective ion size of $i^{th}$ ionic species. In this study, potassium and chloride ions are assumed to have identical ionic sizes of 0.6 nm.

The liquid phase is assumed as an incompressible Newtonian fluid following the Navier–Stokes (NS) equations,

$$-\nabla p + \mu \nabla^2 \mathbf{u} - F \sum_{i=1}^{2} z_i c_i \nabla \phi = \mathbf{0} \qquad (5)$$

Here, $p$, and $\mu$ are the hydrodynamic pressure, and dynamic fluid viscosity, respectively. All simulations are carried out with *COMSOL Multiphysics (version 5.6)*. Other details such as boundary conditions are similar as reported in our previous work.[1]

**Figure S3** shows the ratio, $G_{DNP}/G_{2D}$, as a function of the SiN pore diameter, where $G_{DNP}$ is the overall conductance of DNP and $G_{2D}$ is the conductance of the single 2D MoS$_2$ pore. The result obtained from the numerical model (symbols) shows good agreement with the result based on the resistor model (solid line), validating accuracy of the modified PNP-NS model. To compare the fractional blockade of the DNP device with that of a single pore at the same applied voltage of 400 mV in both cases, we also modeled the current blockade of amino acid in a single nanopore. The 2D pore is 1.1 nm in diameter and 1 nm thick. The SiN pore is 3 nm in diameter and 3 nm thick. The calculated fractional blockades are ~ 20 % and 3 % for the case of 2D and SiN pores, respectively.

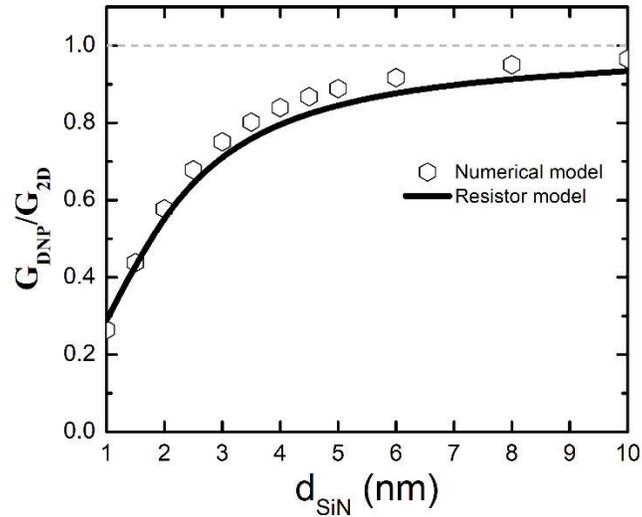

**Figure S3.** Normalized conductance of the DNP as a function of the SiN pore diameter, $d_{SiN}$. For 2D pore, diameter is 1.1 nm and thickness is 1 nm. The thickness of the SiN pore is fixed to 3 nm.

Open symbols represents the results based on the numerical model; solid curve represents the result based on a. simple resistor model (**Section 5**) .

### 7.  Ionic Measurements and Device Wetting Procedure

The wetting procedure was developed for this work. Right after drilling the nanopore in the 2D layer, the device was immersed in the pure EtOH solution. Before the measurement, the device was soaked in the degassed EtOH : $H_2O$ (v/v = 1/1) solution for 1 hour. Customized PMMA flow cells were used where the devices were sealed with silicone gaskets. Two flow cells were filled with 1M KCl solution, which was prepared with DI water. Elements 10 MHz amplifier was utilized to record the current traces by applying an external bias voltage *via* a two-terminal set of Ag/AgCl electrodes. The sampling rate was set as 250 kHz. I-V scans were conducted with voltages up to 500 mV to assess pore wetting by monitoring the resulting current. If the device was unwet, I-V scans were continuously performed for 2 to 3 hours. Alternatively, EtOH/DI mixture was introduced into one of the flow cells to assist in pore wetting. If the pore remained unwet, it was stored in EtOH : $H_2O$ (v/v = 1/1) overnight and we continued with the measurements the next day. For only a few devices on which significant carbon contamination was observed during STEM drilling, an rapid thermal annealing (RTA) process was applied before soaking them in the EtOH/DI solution.

# References

(1) Chou, Y.-C.; Lin, C.-Y.; Castan, A.; Chen, J.; Keneipp, R.; Yasini, P.; Monos, D.; Drndić, M. Coupled Nanopores for Single-Molecule Detection. *Nat. Nanotechnol.* **2024**, *19* (11), 1686–1692. https://doi.org/10.1038/s41565-024-01746-7.

(2) Kim, M. J.; Wanunu, M.; Bell, D. C.; Meller, A. Rapid Fabrication of Uniformly Sized Nanopores and Nanopore Arrays for Parallel DNA Analysis. *Advanced Materials* **2006**, *18* (23), 3149–3153. https://doi.org/10.1002/adma.200601191.

(3) Kim, M. J.; Wanunu, M.; Bell, D. C.; Meller, A. Rapid Fabrication of Uniformly Sized Nanopores and Nanopore Arrays for Parallel DNA Analysis. *Advanced Materials* **2006**, *18* (23), 3149–3153. https://doi.org/10.1002/adma.200601191.

(4) Rodríguez-Manzo, J. A.; Puster, M.; Nicolaï, A.; Meunier, V.; Drndić, M. DNA Translocation in Nanometer Thick Silicon Nanopores. *ACS Nano* **2015**, *9* (6), 6555–6564. https://doi.org/10.1021/acsnano.5b02531.

(5) Liu, K.; Feng, J.; Kis, A.; Radenovic, A. Atomically Thin Molybdenum Disulfide Nanopores with High Sensitivity for DNA Translocation. *ACS Nano* **2014**, *8* (3), 2504–2511. https://doi.org/10.1021/nn406102h.

(6) Hoogerheide, D. P.; Garaj, S.; Golovchenko, J. A. Probing Surface Charge Fluctuations with Solid-State Nanopores. *Phys. Rev. Lett.* **2009**, *102* (25), 256804. https://doi.org/10.1103/PhysRevLett.102.256804.

(7) Ho, C.; Qiao, R.; Heng, J. B.; Chatterjee, A.; Timp, R. J.; Aluru, N. R.; Timp, G. Electrolytic Transport through a Synthetic Nanometer-Diameter Pore. *Proc. Natl. Acad. Sci. U.S.A.* **2005**, *102* (30), 10445–10450. https://doi.org/10.1073/pnas.0500796102.